\documentclass[12pt]{article}
\title{\bf Mirror particles and mirror matter: 50 years of speculation and search}
\author{L.B. Okun \\ ITEP, Moscow, Russia}
\date{}
\begin{document}
\maketitle

\begin{abstract}
This review describes the history of discovery of violation of
spatial parity P, charge conjugation parity C, combined parity CP.
The hypothesis of existence of mirror particles was called upon by
its authors to restore the symmetry between left and right. The
review presents the emergence and evolution of the concepts
``mirror particles'' and ``mirror matter''. It could serve as a
concise guide to the ``mirror-land''. An important part of the
review is the list of about 250 references with their titles.
\end{abstract}

\section{Introduction}

The terms ``mirror particles'', ``mirror matter'' and ``mirror
world'' refer at present to the hypothetical hidden sector of
particles and interactions which compensate the mirror asymmetry
of the weak interactions of ordinary particles. Mirror particles
are considered to be a possible component of the invisible dark
matter. The history of mirror particles is a history of
intertwining of parity violation and parity degeneracy, strict and
broken mirror symmetry, dark matter in the universe, atomic,
nuclear and high energy physics, cosmology and astrophysics.

\section{1950's. Violation of P and C. \\ Conservation of PC.}

In the middle of the 1950's the so called $\theta\tau$ puzzle
became the most challenging problem of elementary particle
physics. At that time the decays $K^+ \to 2\pi$ and $K^+ \to 3\pi$
were assigned to two different mesons $\theta^+$ and $\tau^+$,
having opposite P-parities. But the masses as well as lifetimes of
$\theta^+$ and $\tau^+$ were suspiciously close. Therefore Lee and
Yang put forward the idea of parity degeneracy \cite{1'}. However,
in April of 1956 at the Rochester conference Feynman referring to
Block asked the crucial question: could it be that parity is not
conserved?

Here are a few excerpts from the Proceedings \cite{1''}:

``J.R. Oppenheimer presiding:

There are the five objects $K_{\pi_3}$, $K_{\pi_2}$, $K_{\mu_2}$,
$K_{\mu_3}$, $K_{e_3}$. They have equal, or nearly equal, masses,
and, identical, or apparently identical, lifetimes. One tries to
discover whether in fact one is dealing with five, four, three,
two, or one particle ...''.

``Yang's introductory talk followed:

... the situation is that Dalitz's argument strongly suggests that
it is not likely that $K_{\pi_3}^+ (\equiv \tau^+)$ and
$K_{\pi_2}^+ (\equiv \theta^+)$ are the same particles''.

``Dalitz discussed the $\tau$-$\theta$ problem ... 600 events ...
when plotted on the ``Dalitz diagram'', give a remarkably uniform
distribution ... This would point to a $\tau$-meson of spin-parity
$0^-$ ...''.

``... Feynman brought up a question of Block's:

Could it be that the $\theta$ and $\tau$ are different parity
states of the same particle which has no definite parity, i.e.
that parity is not conserved ...''

``Yang stated that he and Lee looked into this matter without
arriving at any definite conclusions.''

Presumably Feynman meant a special mechanism of parity violation
through the mixing of degenerate scalar and pseudoscalar mesons.

It is interesting that neither Dalitz, nor Michel, who also
participated in the discussion, mentioned the possibility of
parity violation.

A few months later Lee and Yang suggested that parity is not
conserved in weak decays and proposed experiments to search for
pseudoscalar correlations of spin and momentum \cite{3}. (Their
famous paper was received by Physical Review on June 22,
circulated as a preprint, and appeared in the journal on October
1, 1956.) At the end of this paper, in order to save the
left-right symmetry in a more general sense, the existence of
hypothetical right-handed protons, $p_R$, was considered, though
the term ``mirror particles'' was not used and $p_R$ and $p_L$
were assumed to interact ``with the same electromagnetic field and
perhaps the same pion field''.

(Much later I learned that already in 1952 Michel \cite{mich} had
considered parity violating interactions and pseudoscalar
correlations between momenta of several particles in multiparticle
processes. Wick, Wightman and Wigner considered pseudoscalar
amplitudes \cite{4a}. Purcell and Ramsey suggested \cite{4b} to
test experimentally parity conservation by measuring the electric
dipole moment of the neutron. However, they did not realize (as
Landau did realize subsequently) that electric dipole moment
violates the time-reversal invariance as well. Berestetsky and
Pomeranchuk published a note \cite{4c} on the beta-decay of the
neutron, in which they mentioned a remark by Landau that there
exists actually ten four-fermion couplings ``if one allows, in
addition to spinors for pseudo-spinors''.)

As is well known, the experiments proposed by Lee and Yang were
performed half a year later and found large left-right asymmetries
in the $\beta$-decay of $~^{60}{\rm Co}$ \cite{4} and in
$\pi\to\mu\to e$ decays \cite{5,6}.

Before the results of these experiments were published, Ioffe and
Rudik had submitted to ZhETF a manuscript in which they argued
that the existence of short-lived C-even $K_1^0$-meson and
long-lived C-odd $K_2^0$-meson proved that C-parity was conserved
and hence violation of P-parity would mean (due to CPT-theorem)
violation of T-parity (time reversal invariance). This led them to
the conclusion that P-odd, but T-even ${\rm\bf s p}$ asymmetries
are impossible (${\rm\bf s}$ -- spin, ${\rm\bf p}$ -- momentum).

I vividly recall how ITEP theorists discussed these arguments with
Landau after one of the traditional Wednesday ITEP seminars in
November 1956. (At that time the name ITEP did not exist; it was
called TTL -- Thermo-Technical Laboratory.) The discussion took
place in room No.9, where at that time young theorists worked and
where  my desk was.

Landau was absolutely against parity violation because space is
mirror symmetric. This is analogous to conservation of momentum
and angular momentum, because space is homogeneous and isotropic.
Of course the analogy is not complete, because shifts and
rotations are continuous, while reflections are discrete.

[Half a year earlier the Lebedev Institute hosted the first Moscow
conference on elementary particles in which American physicists
participated \cite{77a,11aa}. I recall that Landau laughed at
Gell-Mann (the youngest of the Americans, but already very
famous), when the latter during his seminar at the Institute of
Physical Problems mentioned that parity violation could be one of
the solutions of $\theta\tau$-problem.\footnote{Gell-Mann repeated
the talk he gave at Landau seminar, also in the office of Tamm. I
was taking notes of both of them. He stopped for a moment and
asked me with a smile: ``What happens if you find at home that the
two records contradict each other?''

In the 1980's Telegdi published very interesting reports on the
history of parity violation \cite{7b,7c}. In \cite{7c} he wrote:
``Murray Gell-Mann emphasized to me ..., that I.S. Shapiro most
strenuously objected to the parity violation idea when M.G.M.
presented the latter in 1956 in the Landau seminar as one of the
possible solutions to the $\tau$-$\theta$ puzzle.''

As I have already mentioned, I remember the objections by Landau,
but I don't recall that they were raised also by Shapiro at the
same seminar.}

Roughly at the same time Landau reacted similarly at an
unpublished note by Shapiro. In this note a Wu-type experiment was
suggested. I learned about it three years later, when Shapiro
moved from the Moscow university to ITEP and showed me his
unpublished note. I remember that there was a wrong statement in
this note: the value of energy is different in left- and
right-handed coordinates, if P is violated.\footnote{Further
exposition of this statement is contained in \cite{111a}.} Later
Shapiro gave this note to the director of ITEP Alikhanov, and it
was lost. There was no copying machine at ITEP.]

But let us return to the discussion in room No.9. During the
discussion I pointed out that the short- and long-lived kaons
might exist not due to C-invariance, as was originally proposed by
Gell-Mann and Pais \cite{11}, but due to even approximate
T-invariance. In that case ${\rm\bf s p}$ asymmetries would be
allowed as well as the decay of long-lived neutral kaon into
$3\pi^0$. As a consequence of this discussion, Ioffe and Rudik
urged me to become a coauthor of their paper with my radical
amendments.\footnote{The scheme in which P and T are violated, but
C is conserved, was at length discussed in \cite{115a} even after
it was falsified by experiment.} At first I refused, but conceded
after Ioffe literally went down on one knee in front of me. Our
article \cite{7} was noticed by Yang and Lee, who together with
Oehme \cite{8} independently but later came to the same
conclusions (see references to \cite{7} in their Nobel lectures
\cite{10,9}).

Another consequence of the discussion was that Landau suddenly
changed his mind and put forward the idea of strict
CP-conservation \cite{12}. At the end of this paper he wrote: ``I
would like to express my deep appreciation to L. Okun, B. Ioffe
and A. Rudik for discussions from which the idea of this paper
emerged''. According to his idea, reflection in a mirror of a
process with particles shows us a non-existent process which
becomes physical only after changing particles into corresponding
antiparticles\footnote{See also \cite{13a}.}.

An excellent example of CP-conjugated particles was presented by
Landau \cite{13} in his theory of massless longitudinally
polarized neutrinos: the spin of $\nu$ is oriented opposite to its
momentum, while spin of $\bar\nu$ is oriented along its momentum;
in other words $\nu$ is left-handed, $\bar\nu$ is right-handed.
Both articles \cite{12} and \cite{13} were published as one
article in Nuclear Physics \cite{14}. Longitudinal neutrinos were
independently considered by Salam \cite{15} and by Lee and Yang
\cite{16}. The longitudinal neutrinos lighted up the road to the
$V-A$ theory \cite{17,18}. According to this theory, in the
relativistic limit ($v/c\to 1$) all elementary fermions become
left-handed in weak interactions of charged currents, while their
antiparticles become right-handed.  Only a few years ago the
discovery of neutrino oscillations made it clear that neutrinos
are not massless and hence the theory of longitudinal neutrinos is
valid only approximately, although in many cases with extremely
high accuracy.

It is worth mentioning here the idea of the baryonic photon
coupled to baryonic charge \cite{ly55}. This article became an
inspiration for further search of leptonic photons, paraphotons
and mirror photons (see below).

\section{1960's. CP-violation}

I liked very much the idea of the strict CP-conservation. But, on
the other hand, I could not understand why the coefficients in the
Lagrangian cannot be complex. Thus, in the lectures at ITEP
\cite{19a} weak interactions of hadrons were described on the
basis of  a composite model assuming CP-conservation. In lectures
in Dubna \cite{19b} and in the book \cite{19c} I insisted that the
experimental test of CP-invariance is   one of the highest
priorities.

A group of Dubna experimentalists led by Okonov searched for
CP-violating decay $K_2^0 \to \pi^+ \pi^-$. They have not found
two-body decays among 600 three-body decays \cite{19d}.
Unfortunately at this stage they were stopped by their lab.
director. The group was unlucky. Two years later a few dozens of
these decays with  the  value of the branching ratio almost
reached in \cite{19d} were discovered by the Princeton group
\cite{2}.

The discovery of the decay $K_2 \to 2\pi$ by Christenson et al.
\cite{2} put an end to Landau's idea of strict CP-conservation
according to which antiparticles look exactly like mirror images
of particles. To avoid this conclusion Nishijima and Saffouri
\cite{2'} put forward the hypothesis of ``shadow universe'' to
explain the two pion decays without CP-violation. According to
\cite{2'}, the decays to two pions observed in 1964  were decays
not of CP-odd $K_2^0$ but of a new hypothetical long-lived CP-even
``shadow'' $K_1^{0\prime}$-meson through its transition into
ordinary $K_1^0$. However, as was shown in \cite{okun}, this
mechanism contradicts the results of the  neutrino experiment,
because shadow $K_1^{0\prime}$-mesons would penetrate through the
shielding and decay into two pions in the neutrino detector, while
such events were not observed.

In the next paper  \cite{1} Kobzarev,  Pomeranchuk, and myself
postulated CPA-symmetry (A -- from Alice) and the existence of
hypothetical mirror particles and of a mirror world. (The modern
terminology, in which mirror matter refers only to duplication of
all our particles (not some of them) was {\it in statu nascendi},
therefore the ``mirror world'' and ``mirror particles'' were used
in \cite{1} practically as synonyms. Note that the Standard Model
did not exist at that time.) According to \cite{1}, mirror
particles cannot participate in ordinary strong and
electromagnetic interactions with ordinary particles. In this
respect they differ from the right-handed protons considered by
Lee and Yang \cite{3}. The hidden mirror sector must have its own
strong and electromagnetic interactions. This means that mirror
particles, like ordinary ones, must form mirror atoms, molecules
and, under favorable conditions, invisible mirror stars, planets
and even mirror life. Moreover, this invisible mirror world can
coexist with our world in the same space.

I recall a weekend hike with Igor Kobzarev in a forest near
Moscow, when I suddenly ``saw'' an invisible mirror train silently
crossing a clearing. It was argued in our paper \cite{1} that such
a situation is impossible. A mirror train needs a mirror globe,
but a mirror globe would gravitationally perturb the trajectory of
our globe. Gravitational coupling between two worlds seemed
indispensable\footnote{We did not know the pioneering articles on
dark matter by Oort \cite{23A} and Zwicky \cite{23C,23B}.}. The
coupling of the two worlds via neutral kaons was considered in
\cite{4'}.

 Mirror particles were discussed at the fourth European
conference on elementary particles (Heidelberg, September 1967)
\cite{24A} and at the Moscow conference on CP-violation (January
1968) (see \cite{24B}).

Perhaps it is worth mentioning a few papers which at first sight
have no direct relation to mirror matter. In \cite{24C}  the
muonic photon was considered and the  transitions between it and
ordinary photon through a muonic loop. This gives a coupling
$\epsilon F_{\alpha\beta} F^\prime_{\alpha\beta}$, where $F$ is
our field. $F^\prime$ -- the new one, while $\epsilon$ is a
dimensionless constant. Because of this coupling, which is at
present called ``kinetic mixing'', muonic neutrino acquires a tiny
electric charge $\epsilon e$ (see also \cite{7'},
\cite{glash1}-\cite{holdom1}, \cite{97f}-\cite{2n},
\cite{141a}-\cite{141f}).

In \cite{44aa}  the gravitational dipole moment of the  proton was
discussed and was shown to be forbidden in the framework of
general relativity. Gravitational interaction of the so called
sterile neutrino was considered in \cite{44bb}.

As is well known, in the mainstream of particle physics, the
quarks and the electroweak theory with spontaneous symmetry
breaking were suggested in  the 1960s. An important note by
Sakharov \cite{43aa} was published, which connected CP-violation
with the baryon asymmetry of the universe and hence with our
existence.

\section{1970's. ``Minimum''. Exotic vacua}

In the 1970's charm, beauty and $\tau$-lepton were discovered and
QCD was formulated, but there was a minimum of publications on
mirror particles. I know only one paper, by Pav\v{s}i\v{c}
\cite{3'}. A relation between mirror symmetry and the structure of
a particle was attempted in it: mirror nucleons are
unconditionally necessary, while mirror leptons are necessary only
if they have internal structure. This differs from the standard
concept of the mirror matter. In 2001 the paper \cite{3'} was
posted on electronic archive with a note: ``An early proposal of
`mirror matter' published in 1974'' \cite{pav}.

Also in the 1970's spontaneous breaking of gauge symmetries was
brought to the cosmological model of hot universe
\cite{25a}-\cite{25c} and the first articles were published on
spontaneous breaking of CP-symmetry \cite{lee}, on domain
structure of vacuum \cite{zeld,zeld2} and on metastable vacuum
\cite{voloshin}. Vacuum domains are a consequence of spontaneous
breaking of discrete symmetry. They appear during cooling of the
universe after the big bang. Thus space itself could be not mirror
symmetric (recall Landau's arguments). Metastable vacuum was
dubbed false vacuum three years later, see \cite{fram}-\cite{cal}.

\section{1980's. Revival}

A revival of interest in mirror particles occurred in the 1980's.
In papers \cite{5'}-\cite{11'} various aspects of hidden sector of
particles and interactions were considered. In \cite{5'} the
existence of new long range forces and of $x$- and $y$-particles
was suggested. According to \cite{5'}, $y$-particles have no
direct interactions with ordinary ones, while $x$-particles serve
as connectors: they have interactions with both ordinary and
$y$-particles. In \cite{6',66'} the existence of gluon-like
$\theta$-bosons was proposed. The role of $\theta$-bosons in the
early universe was discussed in \cite{dolg}. They have large
confinement radius and can form unbreakable strings with length
measured in kilometers. In \cite{7'} mirror hadrons and neutral
meson connectors were discussed. The existence of hidden
paraphotons was suggested in \cite{9'}. The mixing of paraphotons
leads to photon oscillations discussed in \cite{9'}. Tiny charges
of  particles which are usually considered to be neutral (such as
atoms and neutrinos) were discussed in \cite{okun1}. A review of
hypothetical phenomena was presented in the rapporteur talk
``Beyond the Standard Model'' \cite{50B}. It contained, among
other subjects, photon oscillations and left-right symmetric
models, but no mirror particles.

In 1986 Ellis visited ITEP and suggested to write a review about
``nothing''. Together with Voloshin we wrote the review
\cite{11'}, a part of which was dedicated to mirror particles.
However at the last moment I decided not to submit it to Soviet
review journal Uspekhi Fiz. Nauk as a too speculative one,
therefore it was published only as an ITEP preprint.

Voloshin continued the quest for mirror particles. He induced the
ARGUS collaboration at DESY to search for decays $\Upsilon(2S) \to
\pi^+ \pi^- \Upsilon(1S)$, in which $\Upsilon(1S)$ due to
transitions to its mirror counterpart decays into ``nothing''. The
upper limit for the branching ratio of this invisible channel was
established: 2.3\% at the 90\% CL \cite{51a,51b}. The search for
invisible decay products of $\phi$-meson was carried out in
\cite{0311}.

An experimental group at ITEP measured the spectrum of electrons
in the tritium $\beta$-decay and announced that the mass of the
electron neutrino is 30 eV. This prompted Zeldovich and Khlopov to
publish their review \cite{zel}. Alongside with other scenarios
they discussed mirror neutrinos and their implications for
cosmology and astrophysics. The neutrino mass was considered by
them as a bridge to the mirror world, the transitions between our
left-handed neutrino and mirror right-handed neutrino being
responsible for $m_\nu$.

A model in which connectors between  the two worlds are the so
called hybrids (particles with electroweak quantum numbers of
ordinary quarks, and mirror QCD quantum numbers of mirror quarks
and their mirror counterparts) was considered in \cite{53B}.
  This model contains bound states with fractional charges -- fractons.

Schwarz and Tyupkin \cite{schwarz} suggested unification of mirror
and ordinary particles within an SO(20)-symmetry. In this model,
mirror cosmic strings appeared. After circling such an ``Alice
string'' an ordinary particle is transformed into a mirror
particle and vice versa (see also \cite{sch} and \cite{0244}).

Further cosmological and astrophysical manifestations of mirror
particles were discussed in \cite{8'}-\cite{goldberg}. In
particular Alice strings were considered in \cite{8',53A}.

Glashow and his coworkers became interested in  the mirror
universe and photon oscillations \cite{glash1,glash2,carlson}. In
\cite{glash2} he suggested to explain experimental anomaly in
orthopositronium decays, observed at that time, by transitions
between orthopositronium and mirror orthopositronium. (For later
developments of this suggestion see \cite{106}, \cite{35aa} -
\cite{2n}.)

Tiny ordinary electric charges of mirror particles which otherwise
have no ordinary charge, but have a mirror electric charge, appear
due to mixing of ordinary and mirror photons \cite{holdom,holdom1}
(see also \cite{0196}).

\section{1991--2006. ``Maximum''. From cosmology and astrophysics
to LHC}

A flood of mirror particle articles occurred after 1990.

The Australian physicist Foot became a great enthusiast of mirror
particles and published dozens of articles  on this subject. One
can appreciate the range of his interests by looking at the titles
of references \cite{1111} - \cite{400} and of his book \cite{122}.
In \cite{1111} mirror symmetrical version of the standard gauge
model was considered, in particular renormalizable mixing
interaction of ordinary and mirror higgses was analyzed. Let us
note that in this renormalizable model the strong transitions of
three quarks into three mirror quarks or of two gluons into two
mirror gluons are forbidden because such interactions are
non-renormalizable. Therefore the discovery of a transition of
neutron into a mirror neutron $n\leftrightarrow n^\prime$ or of
$\Upsilon \leftrightarrow \Upsilon^\prime$ would falsify this
theoretical model.

Let us also mention the new fantastic idea of grains of mirror
matter embedded in the ordinary matter due to the interaction
caused by mixing of ordinary and mirror photons
\cite{125,130,127}. Many coauthors of Foot -- Volkas, Ignatiev,
Mitra, Gninenko, Silagadze -- published also their own papers on
mirror particles \cite{97a}-\cite{0140} (see also \cite{0309}). A
few of these papers were also devoted to mirror grains
\cite{97e,97f}, \cite{sil1}-\cite{0140}.

An impressive contribution to the field of mirror particles
belongs to Berezhiani, who together with his coauthors  published
over 15 papers on various subjects in mirror physics, mirror
astrophysics and mirror cosmology \cite{5a}-\cite{3a} (see also
\cite{17a}-\cite{015}).

Most of the papers cited in this section are based on strict
mirror symmetry. They ascribe the observed macroscopic disparity
between mirror and ordinary particles to the inflation stage of
the universe (see \cite{11a,10n}).

Mohapatra  published about 15 papers (many of them with coauthors)
on various aspects of  mirror astrophysics \cite{moh1}-\cite{0307}
in the framework of broken mirror symmetry.

The search \cite{98b}-\cite{102b} for gravitational microlenses
produced by separate stars in the halos  of galaxies -- the
so-called MACHOS (MAssive Compact Halo Objects) -- has led to the
discovery of an excess of machos in the direction of the Large
Magellanic Cloud \cite{101b,102b}. Even before this discovery,
theorists had indicated \cite{hod,sil,14a} that some of the machos
could be mirror stars. This interpretation was developed further
in \cite{103b,105b}. Though the discovery of machos
\cite{101b,102b} was questioned \cite{106b,107b} (see the
discussion \cite{108b,109b}), many astrophysicists believe that
the stellar dark matter cannot consist of ordinary baryons
\cite{110b,111b}.

Since the times of Oort \cite{23A} and Zwicky \cite{23C,23B} there
existed two alternative explanations of anomalously high
velocities of stars and galaxies (the so called ``virial
paradox''): 1) invisible dark matter, 2) anomalously strong
gravitational force at large distances. Recent observations
\cite{149A,149B} of colliding clusters of galaxies seem to settle
the ambiguity in favor of dark matter. The dark matter which
manifests itself through the effect of gravitational lensing, is
definitely segregated from the luminous parts of the clusters. If
this dark matter is mirror matter the mirror stars in it must be
more prominent compared to mirror gas than ordinary stars compared
to ordinary gas (Blinnikov, Silagadze, private communications).

The correlation of gamma ray bursts with the distribution of dark
matter in the galaxies might suggest that these bursts are
produced by mirror stars either via mirror neutrinos \cite{A} -
\cite{C}, or via mirror axions \cite{1157a,7a,016} (see also
\cite{015}).

For Supernova constraints on sterile neutrino production see
\cite{D}.

Cosmic mirror strings as sources of cosmic rays of ultra-high
energies were considered in \cite{132A} (see also
\cite{07}-\cite{0308B}). Various aspects of mirror astrophysics
were discussed in articles \cite{5n,155b} and books \cite{6n,7n}.
New gauge mirror-type symmetry SU(2) was proposed in \cite{136n}
and analyzed in \cite{137n}-\cite{141}. For discussions of
leptonic (muonic) photons in the 1990's see
\cite{141a}-\cite{141f}. Upper limits on invisible decays of
$B^0$-meson, $\eta$- and $\eta^\prime$-mesons were established in
\cite{51c,51d}. With the Belle detector the upper limit $ 2.5\cdot
10^{-3}$  for the branching fraction $B(\Upsilon (1S) \to $
invisible) was established \cite{238a}.

For sundry ``mirror matters'' see \cite{0196}-\cite{0310}; for
proposed searches of dark matter see \cite{2401}-\cite{240c}.

The first physical run of a special position accumulator ring
LEPTA was conducted in 2004 in Dubna. One of the aims of this ring
is the search for mirror orthopositronium
\cite{mesh1}-\cite{mesh6}.

A very interesting discussion of invisible decay channels of
higgses, which can be searched for at Large Hadron Collider, can
be found in  \cite{8n}, \cite{142}-\cite{144} (see also
\cite{207}). The invisible decays appear due to the mixing of the
ordinary and mirror higgses. The higgs could be discovered in the
near future.

\section{Concluding remarks}

One might introduce an acronym for mirror symmetry: mirsy,
analogous to susy for supersymmetry. Compare mirsy with susy.
Mirsy cannot compete with susy in  the depth of its concepts and
mathematics. But I believe it can compete in the breadth and
diversity of its phenomenological predictions. Certainly, mirror
matter is richer than the dark matter of susy.

The preliminary version of this review was prepared for the talk
at the ITEP Meeting on the Future of Heavy Flavor Physics, July
24-25, 2006 (http://www.itep.ru/eng/bellemeeting) and published on
June 19 as hep-ph/0606202 v.1. The final version (v.2) was
prepared for the Russian review journal Uspekhi Fiz. Nauk during
the summer of 2006.

As a result the number of references has doubled. It could have
risen even higher. If you type in Google ``mirror particles'' (do
not forget quotation marks!), you get a thousand entries (sites).
(If you type ``mirror world'', or ``mirror universe'', you get
about 200000 entries devoted mainly to Star Trek television
episodes.)  A search in Wikipedia is suggested in some of the
entries. But the Wikipedia articles on mirror matter are rather
misleading. Instead of Google it is better to use Google Scholar,
where the number of entries for ``mirror universe'' is about a
hundred, while for ``mirror particle'' -- a few hundred. The extra
articles in Google Scholar do not deal with those mirror particles
that are the subject of this review. They are ``mirror'' in a
different sense. For instance, the terms ``mirror families'' or
``mirror fermions'' refer to hypothetical families of very heavy
fermions with reversed isotopic quantum numbers, which are
presumed to interact with ordinary photons and gluons.

\section{Acknowledgements}

I am grateful to M.V. Danilov for inviting me to give a talk on
mirror matter at the ITEP meeting. I would like to thank Z.G.
Berezhiani, S.I. Blinnikov, O.D. Dalkarov, A.D. Dolgov, S.N.
Gninenko, A.Yu. Ignatiev, M.Yu. Khlopov, Z.K. Silagadze, R.R.
Volkas and M.B. Voloshin for very valuable suggestions,  as well
as T. Basaglia, E.A. Ilyina and O.V. Milyaeva for their help in
preparing this review. This work was partly supported by grant
NSh-5603.2006.2.

\end{document}